\documentclass[a4paper]{article}

\usepackage{INTERSPEECH2019}

\usepackage{epsfig,amssymb,amsmath}
\usepackage{cite}
\usepackage{acronym}
\usepackage{multirow,makecell} 
\usepackage{todonotes}
\usepackage{url}

\newacro{TTS}{Text-to-Speech}
\newacro{M-TTS}{Multi-speaker Text-To-Speech}
\newacro{LDE}{Learnable Dictionary Encoding}
\newacro{LDA}{Linear Discriminant Analysis}
\newacro{ASR}{Automatic Speech Recognizer}
\newacro{WSJ}{Wall Street Journal}
\newacro{WER}{Word-Error-Rate}
\newacro{CER}{Character-Error-Rate}
\newacro{EER}{Equal-Error-Rate}
\newacro{NN}{Neural Network}
\newacro{ResNet}{Residual Network}
\newacro{SR}{Sub-sampling Rate}
\newacro{SV}{Speaker Verification}
\newacro{TDNN}{Time Delay Neural Network}

\title{Learning Speaker Embedding from Text-to-Speech}

\name{Jaejin Cho$^1$, Piotr \.Zelasko$^1$, Jes\'us Villalba$^{1,2}$, Shinji Watanabe$^{1,2}$, Najim Dehak$^{1,2}$}

\address{$^1$Center for Language and Speech Processing,
  $^2$Human Language Technology Center of Excellence,  \\ Johns Hopkins University, Baltimore, MD, USA}
\email{\{jcho52,pzelask2,jvillal7,shinjiw,ndehak3\}@jhu.edu}
\begin{document}

\maketitle
\begin{abstract}
Zero-shot multi-speaker Text-to-Speech (TTS) generates target speaker voices given an input text and the corresponding speaker embedding. 
In this work, we investigate the effectiveness of the TTS reconstruction objective to improve representation learning for speaker verification. 
We jointly trained end-to-end Tacotron 2 TTS and speaker embedding networks in a self-supervised fashion. We hypothesize that the embeddings will contain minimal
phonetic information since the TTS decoder will obtain that information from the textual input. TTS reconstruction can also be combined with speaker classification to enhance these embeddings further. Once trained, 
the speaker encoder computes representations for the speaker verification task, while the rest of the TTS blocks are discarded. We investigated training TTS from either manual or ASR-generated transcripts. The latter allows us
to train embeddings on datasets without manual transcripts. 
We compared ASR transcripts and Kaldi phone alignments as TTS inputs, showing that the latter performed better due to their finer resolution. 
Unsupervised TTS embeddings improved EER  by 2.06\% absolute with regard to i-vectors for the LibriTTS dataset. 
TTS with speaker classification loss improved EER by 0.28\% and 0.73\% absolutely from a model using only speaker classification loss in LibriTTS and Voxceleb1 respectively.
\end{abstract}

\section{Introduction}

Neural \ac{TTS}~\cite{shen2018taco2} is gaining great attention due to its simpler system pipeline and improved performance compared to a more conventional statistical \ac{TTS} system \cite{zen2009statistical}. 
Several neural \ac{TTS} models include speech encoder modules that aim to extract latent representation to control desired characteristics such as speaker voice, accent, speaking style, or noise to the synthesized speech~\cite{nachmani2018fittingnewspk_fb,ping2017DV3,jia2018transfer,wang2018styletoken,Hsu2019DisentSpknNoise}, in addition to a text encoder.

Among those models, \ac{M-TTS} model can synthesize speech imitating the voices of multiple speakers. A key component in \ac{M-TTS} systems is the speaker encoder that extracts a speaker embedding from one or several utterances of the speaker of interests. This embedding is used to customize the TTS output and generate new utterances from the target speaker. The speaker encoder can be jointly trained with the rest modules of the \ac{M-TTS}. The speaker encoder can also be trained ahead and used to produce embeddings
to train \ac{M-TTS} on a multi-speaker dataset~\cite{jia2018transfer}.
While most of the previous papers focused on improving \ac{TTS} as their final goal, this work's core focus is analyzing the utility of the speaker encoder for \ac{SV} tasks.

There are three main reasons we expect the \ac{TTS} to help learn a better speaker embedding.
Firstly, by leveraging the encoded latent representation from the reference speech, \ac{TTS} controls several aspects of the synthesized speech, such as a speaker's voice, speaking style/prosody, and noise. Thus, training an \ac{M-TTS} that naturally synthesizes desired speech of multiple voices, implies that the speaker encoder extracts a robust speaker embedding, sufficient to discriminate between the voices of different speakers. Building controllable robust \ac{TTS} systems has been the aim of several past works, as in~\cite{hsu2018hierarchicaltts,henter2018s2sContTTS,akuzawa2018TTSwVAE}.
In our paper, we focus on improving the speaker encoder module. To that end, we use a system proposed previously in~\cite{jia2018transfer}, with a major difference -- the speaker encoder is trained jointly with the rest of \ac{TTS} modules~\cite{nachmani2018fittingnewspk_fb}.

Secondly, it has been previously observed that speaker recognition systems suffer from imbalanced or incomplete coverage of the phonetic variability in the speaker embedding space, especially in short utterances~\cite{vogt2009within}. However, we expect that the speaker encoder in \ac{M-TTS} can amend this problem. We hypothesize that the speaker encoder learns a speaker representation robust to the phonetic variability, given that the phonetic information is mainly encoded by the text encoder. 

Finally, \ac{M-TTS} training does not explicitly require speaker labels but only the paired speech and transcript data.
Also, if we even do not have transcripts, a well pre-trained \ac{ASR} can generate pseudo labels for the speech-only data.
Thus, it enables unsupervised learning of a speaker embedding similarly as i-vector, the state-of-the-art unsupervised method for learning a speaker embedding~\cite{dehak2010ivector}.
Note that this setup is similar to recent activities in self-supervised speaker embeddings~\cite{stafylakis2019ssspkemb}, unsupervised/self-supervised speech processing including ASR/TTS joint modeling~\cite{tjandra2017listening,hori2019cycle}, voice conversion~\cite{van2017neural,lorenzo2018voice}, and zero speech challenge (TTS without T)~\cite{dunbar2019zero}.
However, again, the main difference between them and our work is that the primary focus in this paper is to investigate the effectiveness of the learned speaker embedding for the \ac{SV} task in this unsupervised setup.


In our experiments, we first compared different types of input texts in \ac{TTS} training, including a human transcript, an \ac{ASR}-generated transcript, and phone alignment. 
We found that replacing the human transcript with \ac{ASR}-generated outputs does not degrade the speaker encoder training in \ac{TTS}. Also, using the phone alignment input works better than using the \ac{ASR} transcript in training the \ac{TTS} for the speaker encoder. We compared the proposed method with other state-of-the-art methods and found that it consistently outperforms them in most experimental setups, both in supervised and unsupervised settings. We conclude that learning the speaker embedding with the \ac{TTS} criterion indeed helps with the \ac{SV} task.
   
\section{Learning Speaker Embedding with Text-to-Speech}
The main goal of this paper is to investigate whether an \ac{M-TTS} model can help the speaker encoder to learn better embeddings for the \ac{SV} task. To that end, we first trained an \ac{M-TTS} based on the Tacotron 2~\cite{shen2018taco2} architectural design. The original Tacotron 2 is composed of a text encoder, a decoder, and a vocoder. 
A text encoder $\mathrm{Enc}^{\mathsf{txt}}(\cdot)$ encodes a $J$-length text input sequence $W=\{w_i \in \mathcal{V}|i=1, ..., J\}$ with a vocabulary $\mathcal{V}$ into a sequence of $D$-dimensional hidden vectors $\mathbf{H}=\{\mathbf{h}_i \in \mathbb{R}^D|i=1, ..., J\}$ as follows:
\begin{equation}
    \mathbf{H} = \mathrm{Enc}^{\mathsf{txt}}(W).
    \label{eq:enc_txt}
\end{equation}
Then, the decoder $\mathrm{Dec} (\cdot)$ predicts a $T$-length sequence of target acoustic features $\mathbf{O} = \{\mathbf{o}_t \in \mathbb{R}^{D_{\mathsf{a}}}| t=1, ..., T\}$ with $D_{\mathsf{a}}$-dimensional features, e.g., Mel-filter-banks, based on a context vector generated by an attention mechanism over hidden vectors $\mathbf{H}$ as follows.
\begin{equation}
    \mathbf{o}_t = \mathrm{Dec} (\mathbf{o}_{t-1}, \mathbf{H})
    \label{eq:dec_txt}
\end{equation}
The prediction happens in an auto-regressive fashion, i.e., the prediction of $\mathbf{o}_t$ is performed with the previous acoustic feature $\mathbf{o}_{t-1}$ as a condition.
The condition $\mathbf{o}_{t-1}$ would be a ground truth defined as $\mathbf{o}^{\ast}_{t-1}$ during training (so called teacher-forcing) or a predicted one during inference.

To enable this model to generate  voices of multiple speakers, a speaker encoder $\mathrm{Enc}^{\mathsf{spk}}(\cdot)$ is added to encode a $D_{\mathsf{e}}$-dimensional global speaker embedding vector $\mathbf{e} \in \mathbb{R} ^{D_{\mathsf{e}}}$ as follows:
\begin{equation}
    \mathbf{e} = \mathrm{Enc}^{\mathsf{spk}}(\mathbf{O}).
    \label{eq:emb}
\end{equation}
The speaker embedding vector $\mathbf{e}$ is concatenated to every encoded text vector $\mathbf{H}$ over the sequence~\cite{jia2018transfer}.
That is, the multispeaker decoder function is extended from Eq.~\eqref{eq:dec_txt} as follows:
\begin{equation}
    \mathbf{o}_t = \mathrm{Dec}^{\mathsf{mlt}} (\mathbf{o}_{t-1}, \mathrm{Cat}(\mathbf{H}, \mathbf{e})),   
\end{equation}
where $\mathrm{Cat}(\cdot)$ is a concatenation function between $\mathbf{H}$ and $\mathbf{e}$. 
We exclude the explanation of the vocoder, which is usually trained separately to generate a waveform from the Mel-spectrogram. 
While in~\cite{jia2018transfer}, they use a pretrained 
speaker encoder, we propose to jointly train the speaker encoder $\mathrm{Enc}^{\mathsf{spk}}(\cdot)$ in Eq.~\eqref{eq:emb} with the rest of the TTS blocks.
Thus, the TTS reconstruction loss $\mathcal{L}^{\mathsf{tts}}$ is defined as follows: 
\begin{align}
    \mathcal{L}^{\mathsf{tts}} = \sum _{t} \left |\mathbf{o} ^{\ast}_t - \mathrm{Dec}^{\mathsf{mlt}} (\mathbf{o} ^{\ast}_{t-1}, \mathrm{Cat}(\mathbf{H}, \mathrm{Enc}^{\mathsf{spk}}(\mathbf{O}^{\ast})) \right |_p,
    \label{eq:tts_loss}
\end{align}
where $|\cdot|_p$ denotes an Lp-norm.
The actual TTS loss function is a combination of the L1 and L2 losses.
\ac{M-TTS} training does not require speaker labels to learn a embedding extractor, which can be used to reconstruct the speech from different speakers. 

\begin{figure}
    \centering
    \includegraphics[width=1.0\linewidth]{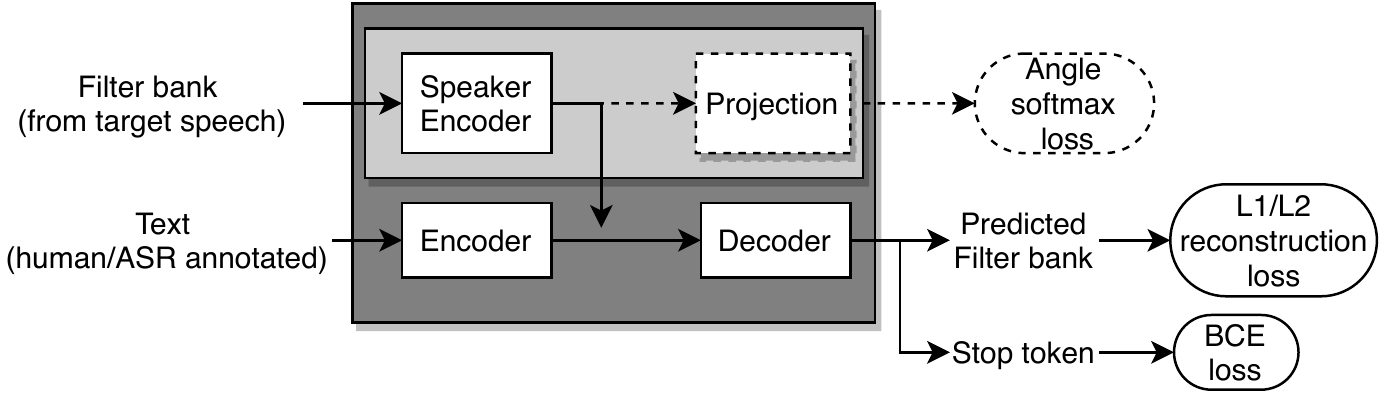}
    \caption{Multi-speaker \ac{TTS} module used to learn speaker embedding. The outer most black box represents the \ac{TTS} module while the grey box inside represents the speaker-related module. Square boxes are modules of the system, and rounds are losses. Dotted parts are added when a speaker classification loss (angle softmax loss here) is added.}
    \label{fig:overview}
      \vspace{-0.15in}
\end{figure}

If there is an available speaker ID label $l^{\ast}_s$ for a speaker $s$, an additional projection layer $\mathrm{Proj}(\cdot)$ can be added to the speaker encoder to calculate the speaker classification loss.
\begin{equation}
    \mathcal{L}^{\mathsf{spk}} = \mathrm{AngSoftMax}(l^{\ast}_s, \mathrm{Proj}(\mathbf{e}_s)),
    \label{eq:spk_loss}
\end{equation}
where $ \mathrm{AngSoftMax}(\cdot)$ denotes an angular softmax loss used for speaker classification~\cite{liu2017angularsfmx}.
$\mathbf{e}_s$ is an embedding vector obtained by Eq.~\eqref{eq:emb} with the acoustic features of the speaker $s$.
This can be considered as multi-task learning for both~\ac{TTS} and speaker classification. 
The diagram for the aforementioned description is in Figure~\ref{fig:overview}. 
In the figure, the \ac{TTS} loss $\mathcal{L}^{\mathsf{tts}}$ is a sum of L1, L2 reconstruction losses for filter bank prediction, as introduced in Eq.~\eqref{eq:tts_loss} and the additional binary cross entropy loss for the stop token prediction. 
Optionally, we can also include the speaker classification loss $\mathcal{L}^{\mathsf{spk}}$, as introduced in Eq.~\eqref{eq:spk_loss}.
Once the \ac{M-TTS} training finishes either with or without speaker classification loss, only the speaker encoder is used to extract embeddings for \ac{SV}.

%

Notably, most of the speech recordings do not include human transcripts $W$ used in Eq.~\eqref{eq:enc_txt}, except for \ac{ASR}-oriented corpora, since obtaining human transcripts is a lengthy and expensive process. 
Therefore, it is important to check whether automatically generated text inputs from \ac{ASR}s are suitable for training \ac{M-TTS} for the speaker encoder. 
For that purpose, we explored using transcripts generated from ASR systems, that is $\hat{W} = \mathrm{ASR}(\mathbf{O})$ instead of $W$ as an input of Eq.~\eqref{eq:enc_txt}.
We can have different ASR systems with different \ac{WER}s, as well as using a phone alignment from a hybrid \ac{ASR} system.

\section{Experimental Setup}
\subsection{Datasets}

We use two datasets in our experiments. The first dataset was the \textit{train-clean-100} and \textit{train-clean-360} subsets of the LibriTTS~\cite{zen2019libritts}, which is read speech corpus designed for \ac{TTS} research. 
These subsets are composed of speech considered clean in terms of signal-to-noise ratio (SNR) since their SNRs are higher than 20dB. 
The subsets were divided into \textit{dev} and \textit{test}, having 1000 and 150 speakers respectively, without speaker overlap. The \textit{dev} was used for training \ac{M-TTS} and \ac{SV} back-end; the \textit{test}  was used for the \ac{SV} evaluation using the extracted embeddings. We created \ac{SV} trials from the \textit{test} set, where each trial is a pair of enrollment and test utterances. We made all possible utterance pair combinations removing cross-gender pairs.
For each trial, the system determines whether both utterances in the trial belong to the same speaker or not. 
This dataset has both human transcript and speaker ID labels. With this dataset, we would like to see how appropriately the \ac{M-TTS} system works depending on the level of supervision, i.e., the amount of the labels used in training, by comparing the systems to an i-vector~\cite{dehak2010ivector} unsupervised system and a \ac{ResNet}-based supervised system~\cite{villalba2018jhumitNISTsre2018}.

To test the proposed systems on a more challenging and closer to the real-world scenario, we also used Voxceleb1 dataset~\cite{nagrani2017voxceleb1}. This corpus is composed of conversational speech utterances with moderate noise, which are processed from interview videos of 1,251 celebrities uploaded on Youtube. The corpus does not have human transcripts but has speaker labels. The Voxceleb1 \textit{dev} and \textit{test} subsets were used for training and evaluating the model, respectively. No data augmentation was done in training.
\subsection{\ac{M-TTS} system configuration}
We used ESPnet-TTS~\cite{hayashi2020espnet} as our \ac{M-TTS} system. We are planning to make this proposed \ac{SV} system publically available as an open-source~\footnote{\url{https://github.com/JaejinCho/espnet_spkidtts.git}}.

For the speaker encoder within the \ac{M-TTS} system, we used the same network design, \ac{ResNet}-LDE, as in~\cite{villalba2018jhumitNISTsre2018}. 
The \ac{ResNet}-LDE network is different from the original x-vector~\cite{snyder2018xvector} system in that it replaces \ac{TDNN} layers with a residual network with 2D convolution layers and replaces the pooling layer with a \ac{LDE} layer~\cite{cai2018LDE}. 

\begin{table}
    \centering
    \caption{EER(\%) in \ac{SV} according to the change of speaker classification loss weight}
    \resizebox{\columnwidth}{!}{%
    \begin{tabular}{cccccccc}
        \toprule
        \multicolumn{2}{c}{Spkloss\_W} & 0 & 0.001 & 0.01 & 0.03 & 0.3 & 3 \\ \hline
        \makecell{EER\\(\%)} & \makecell{LibriTTS\\Voxceleb1} & \makecell{1.36\\9.38} & \makecell{1.31\\8.64} & \makecell{1.16\\\textbf{4.49}} & \makecell{\textbf{1.08}\\4.53} & \makecell{1.13\\4.80} & \makecell{1.20\\5.26} \\
        \bottomrule
    \end{tabular}}
    \label{tab:eer_vs_spkidloss}
    \vspace{-0.1in}
\end{table}
When an additional speaker classification (angular softmax) loss (dotted parts in Figure~\ref{fig:overview}) was added to the \ac{M-TTS} loss, the speaker loss was weighted before added. Table~\ref{tab:eer_vs_spkidloss} shows how the weight value affects performance. 
Note that the weight value 0.03 showed the lowest \ac{EER} on LibriTTS while it showed the 2nd lowest \ac{EER} on Voxceleb1 with almost no difference to the lowest one. Thus, results for the \ac{M-TTS} plus speaker classification loss systems are reported using 0.03 for speaker loss weight throughout the paper.

 For the \ac{SV} back-end system training, the LDA dimension reduction to 150, followed by PLDA~\cite{kenny2010bayesian}, was used throughout all the experiments.
    
\subsection{\ac{ASR} systems description}

To examine how \ac{ASR}-generated text inputs compared to the human transcript affect the \ac{TTS} training for speaker embeddings, three \ac{ASR} models were used. The first \ac{ASR} model was a joint CTC-attention based end-to-end model~\cite{kim2017joint} with convolution and Long Short-Term Memory (LSTM) layers. The training corpus was the \ac{WSJ} corpus~\cite{paul1992wsj}. The second one was also the same end-to-end model but with transformer architecture~\cite{karita2019rnnVSxformerinspeech} and trained on the LibriSpeech corpus~\cite{panayotov2015librispeech}. The third model was a hybrid \ac{ASR}~\cite{povey2011kaldi} model using factorized \ac{TDNN}~\cite{povey2018ftdnn} trained with Lattice-Free Maximum-Mutual-Information (LF-MMI) criterion~\cite{povey2016lfmmi}, also using Librispeech. With this model, we generated both the transcripts and the phonetic alignments.

The \ac{WER}s calculated on the \textit{train-clean-100} and\textit{ train-clean-360} subsets of the LibriTTS were 44.0, 2.7, and 5.15(\%) for the first, second, and third ASR models respectively. The \ac{WER}s could not be calculated on the Voxceleb1 corpus since there was no human transcript available. However, we expected them to be worse compared to the \ac{WER}s on the LibriTTS subsets, considering the Voxceleb1 corpus is more challenging due to mismatched acoustic conditions and spontaneous speaking style. Pre-trained models available online were used for the second end-to-end model and the hybrid model.

\section{Results and Analysis}
%
\begin{table}
\centering
\caption{Comparison between different text inputs for \ac{M-TTS} training: \ac{SV} evaluation results on LibriTTS with \ac{EER}(\%) and MinDCF with \textit{p}=0.01. Human Trans. means human transcript.}
\scalebox{1.03}{
\resizebox{\columnwidth}{!}{%
\begin{tabular}{@{}cccccc@{}}
\toprule
\multirowcell{2}{\ac{TTS} text input\\(ASR training)} & \multirowcell{2}{WER(\%)} & \multicolumn{2}{c}{\ac{M-TTS}} & \multicolumn{2}{c}{\ac{M-TTS} + SpkID loss} \\
& & EER(\%) & MinDCF & EER(\%) & MinDCF\\ \hline
          
Manual Trans. & N/A	& 1.34 & 0.510 & 1.13 & 0.502 \\ \hline
\makecell{E2E ASR Trans.\\(WSJ)} & \makecell{44.00}	& \makecell{1.36} & \makecell{0.510} & \makecell{1.08} & \makecell{0.497} \\
\makecell{E2E ASR Trans.\\(LibriSpeech)} & \makecell{2.70}	& \makecell{1.32} & \makecell{0.511} & \makecell{1.06} & \makecell{0.496} \\
\makecell{Hybrid ASR Trans.\\(LibriSpeech)} & \makecell{5.15}	& \makecell{1.49} & \makecell{0.516} & \makecell{1.06} & \makecell{\textbf{0.493}} \\ \hline
\makecell{Hybrid Phn. Align.\\ SR1 (LibriSpeech)} & \makecell{N/A}	& \makecell{\textbf{1.12}} & \makecell{\textbf{0.501}} & \makecell{\textbf{1.04}} & \makecell{0.502} \\
\makecell{Hybrid Phn. Align.\\ SR3 (LibriSpeech)} & \makecell{N/A}	& \makecell{1.31} & \makecell{0.524} & \makecell{1.11} & \makecell{0.509} \\

\bottomrule
\end{tabular}
}
}
\label{tab:LibriTTS_cmpr_textinput}
\vspace{-0.1in}
\end{table}
\subsection{LibriTTS results}
\subsubsection{Transcript source analysis}

First, different \ac{TTS} systems, either with or without a speaker classification loss, were trained using different text inputs, either transcribed by human annotators or generated from \ac{ASR} systems. Once the \ac{TTS} systems had been trained, speaker encoders were used to extract speaker embeddings to evaluate an \ac{SV} task. Table~\ref{tab:LibriTTS_cmpr_textinput} shows the comparative results with \ac{EER} and MinDCF at p=0.01. Comparing from the first to fourth rows in Table~\ref{tab:LibriTTS_cmpr_textinput}, we observe that regardless of using a manual or \ac{ASR} transcript, the \ac{SV} performance was not affected.

In the table, the \ac{M-TTS} systems trained with phone alignment inputs with the frame \ac{SR} 1 performed the best. Here, frame \ac{SR} means how many acoustic frames were used to predict one phone label. 
For example, the hybrid \ac{ASR} we used generated one phone label every three acoustic frames, i.e., frame \ac{SR} is 3. Thus, to make \ac{SR} 1, we up-sampled each predicted phone label by 3. 
One possible reason for the best systems could be the aligned phoneme inputs to \ac{TTS} reduce the burden for the speaker encoder to include phoneme or pronunciation information since that information can be more easily learned by the text encoder module. This might also happen with the transcript inputs, but the degree might be less.
Another possible reason is that the silence and short pause duration information obtained by the aligned phoneme input could make TTS training more stable.

Note that there was no significant difference in \ac{SV} performance between using an \ac{ASR} with high \ac{WER} or one with low \ac{WER}, as it is shown between E2E \ac{ASR} Transcript (\ac{WSJ}) and E2E ASR Transcript (LibriSpeech) in Table~\ref{tab:LibriTTS_cmpr_textinput}. Considering that the difference in the \ac{WER}s is quite large, this is an interesting observation. For the 44.0\% \ac{WER} ASR-generated transcript, we investigated the quality of the trained \ac{TTS} outputs, and most of them were nonsensical sounds due to the attention not trained well. Nevertheless, it did not seem to affect the quality of speaker representations negatively. One explanation for this result could be that although the \ac{TTS} failed to learn the attention, it still tries to include in the synthesized speech other speech characteristics such as a speaker's voice to reduce the loss. The attention problem was solved with transcripts from \ac{ASR}s having lower \ac{WER}s, synthesizing reasonable speech.

\subsubsection{Model comparison by level of supervision}

Table~\ref{tab:LibriTTS_cmpr_existVSproposed} compares several models with different levels of supervision. We have i-vector and \ac{ResNet}-\ac{LDE} as previously proposed models.
As unsupervised learning for speaker embedding, i-vector and \ac{M-TTS} without speaker ID loss trained with the \ac{ASR} phone alignment (\ac{M-TTS} w/ ASR Phn. Ali. SR1, the fifth row in Table~\ref{tab:LibriTTS_cmpr_existVSproposed}) were compared, showing the latter system outperformed i-vector largely in both \ac{EER} and MinDCF.

\begin{table}
    \centering
    \caption{Comparison between proposed \ac{M-TTS} models to existing models: \ac{SV} results on LibriTTS with \ac{EER}(\%) and MinDCF with \textit{p}=0.01. Human Trans. means human transcript.}
    \scalebox{1.00}{
    \resizebox{\columnwidth}{!}{%
    \begin{tabular}{lcccc}
        \toprule
        \multirow{2}{*}{System} & \multicolumn{2}{c}{Metrics} & \multicolumn{2}{c}{Label usage} \\
        & EER(\%) & MinDCF & Human Trans. & SpkID\\ \hline
         
        i-vector & 3.18 & 0.639 & & \\ \hline
        \ac{ResNet}-\ac{LDE} & 1.32	& \textbf{0.500} & & $\bigcirc$ \\ \hline
        \ac{M-TTS} & 1.34 & 0.510 & $\bigcirc$ & \\ \hline 
        \ac{M-TTS} + SpkID loss & 1.13	& 0.502 & $\bigcirc$ & $\bigcirc$ \\ \hline
        \makecell[l]{\ac{M-TTS}\\w/ ASR Phn. Ali. SR1} & 1.12 & 0.501 & & \\ \hline
        \makecell[l]{\ac{M-TTS} + SpkID loss \\w/ ASR Phn. Ali. SR1}  & \makecell{\textbf{1.04}} & \makecell{0.502} & & \makecell{$\bigcirc$} \\
        \bottomrule
    \end{tabular}
    }
    }
    \label{tab:LibriTTS_cmpr_existVSproposed}
    \vspace{-0.1in}
\end{table}

Next, as a supervised training setup with either a human transcript or speaker ID labels, two \ac{M-TTS} systems (the 3rd and the last rows in Table~\ref{tab:LibriTTS_cmpr_existVSproposed}) were compared to \ac{ResNet}-\ac{LDE}. Here, \ac{ResNet}-\ac{LDE} showed a similar result to the \ac{M-TTS} system trained with only the human transcript (M-TTS), which suggests that a speaker embedding can be learned implicitly by training an \ac{M-TTS} with a transcript. Meanwhile, an \ac{M-TTS} trained on ASR phone alignments with only speaker ID labels (\ac{M-TTS} + SpkID loss w/ ASR Phn. Align. SR1) outperformed both \ac{ResNet}-\ac{LDE} and M-TTS.

Adding speaker classification loss to \ac{M-TTS} in training (4nd row in Table~\ref{tab:LibriTTS_cmpr_existVSproposed}) outperformed the pure discriminative system (ResNet-\ac{LDE}). This implies that multi-task learning for both \ac{TTS} and the speaker classification enables better speaker embedding learning for the \ac{SV} task. 
%
          

%

To sum up, the unsupervised training setup, \ac{M-TTS} w/ ASR Phn. Ali. SR1, outperformed the i-vector and the state-of-the-art \ac{ResNet}-\ac{LDE} that learns speaker embedding in a supervised way. Then, adding a speaker ID loss to the \ac{M-TTS} w/ ASR Phn. Ali. SR1 outperformed the \ac{ResNet}-\ac{LDE} system further.

\subsection{Voxceleb1 results}
\subsubsection{ASR transcript analysis}

All the \ac{ASR}s used on this dataset were the same as used in LibriTTS. Although the performances of \ac{ASR}s cannot be calculated due to unavailable human transcripts, it is expected that the \ac{WER}s are worse compared to ones calculated on LibriTTS due to domain mismatch between the \ac{ASR} training data (read and clean speech) and Voxceleb1 (conversational and moderately noisy speech).

The results are shown in Table~\ref{tab:Voxceleb1_cmpr_textinput}. \ac{TTS} training with Hybrid \ac{ASR} Transcript (LibriSpeech) was skipped here since it did not improve with regard to E2E in the LibriTTS experiments. The performance gaps between using transcripts and phone alignments became larger on Voxceleb1, compared to LibriTTS. Different from what is observed on LibriTTS, Hybrid ASR Phn. Align. SR3 (LibriSpeech) worked better than Hybrid ASR Phn. Align. SR1 (LibriSpeech). This inconsistent result is possibly due to less accurate phone alignment prediction on Voxceleb1.

\subsubsection{Model comparison by level of supervision}

The systems in the Hybrid ASR Phn. Align. SR3 (LibriSpeech) row in Table~\ref{tab:Voxceleb1_cmpr_textinput} were compared to previously published systems. The results are shown in Table~\ref{tab:Voxceleb1_cmpr_existVSproposed}.
In an unsupervised scenario of learning speaker embedding, two i-vector systems and \ac{M-TTS} w/ ASR Phn. Align. SR3. were compared. Although the 2048-GMM i-vector system outperformed the proposed system, the comparison result could be different with more data since the \ac{NN} based systems are known to require more data in training to perform better than classical methods, e.g. i-vector.

As supervised ways of learning speaker embedding,  we compared \ac{ResNet}-\ac{LDE} and \ac{M-TTS} + SpkID loss w/ ASR Phn. Align. SR3. The proposed model outperformed \ac{ResNet}-\ac{LDE} by 15.08\% relatively in \ac{EER}. It also outperformed SincNet+LIM~\cite{ravanelli2018mutualinfo4spkveri}, which learns semi-supervised embedding using mutual information.

%

%

\begin{table}
\centering
\caption{Comparison between different text inputs for \ac{M-TTS} training on Voxceleb1: \ac{SV} evaluation results on Voxceleb1 with \ac{EER}(\%) and MinDCF with \textit{p}=0.01.}
\resizebox{\columnwidth}{!}{%
\begin{tabular}{ccccc}
\toprule
\multirowcell{2}{\ac{TTS} text input\\(ASR training corpus)} & \multicolumn{2}{c}{\ac{M-TTS}} & \multicolumn{2}{c}{\ac{M-TTS} + SpkID loss} \\
& EER(\%) & MinDCF & EER(\%) & MinDCF\\ \hline

\makecell{E2E ASR Transcript\\(WSJ)} & \makecell{9.38} & \makecell{0.690} & \makecell{4.53} & \makecell{0.443} \\
\makecell{E2E ASR Transcript\\(LibriSpeech)} & \makecell{9.78} & \makecell{0.697} & \makecell{4.32} & \makecell{0.443} \\ \hline
%
%
\makecell{Hybrid ASR Phn. Align.\\ SR1 (LibriSpeech)} & \makecell{7.21} & \makecell{0.617} & \makecell{4.53} & \makecell{0.443} \\
\makecell{Hybrid ASR Phn. Align.\\ SR3 (LibriSpeech)} & \textbf{\makecell{6.37}} & \makecell{\textbf{0.556}} & \makecell{\textbf{4.11}} & \makecell{\textbf{0.416}} \\

\bottomrule
\end{tabular}
}
\label{tab:Voxceleb1_cmpr_textinput}
\end{table}
%
%

\begin{table}
    \centering
    \caption{Comparison of models by supervision level on Voxceleb1: \ac{SV} results on Voxceleb1 with \ac{EER}(\%) and MinDCF with \textit{p}=0.01.}
    \resizebox{\columnwidth}{!}{%
    \begin{tabular}{lccc}
        \toprule
        System & EER(\%) & MinDCF & SpkID usage\\ \midrule
        1024-GMM i-vector~\cite{nagrani2017voxceleb1} & 8.8 & n/a & \\ \hline
        2048-GMM i-vector~\cite{peng2020mixture} & 5.51 & 0.462 & \\ \hline
        \ac{ResNet}-\ac{LDE} & 4.84	& 0.484 & $\bigcirc$ \\ \hline
        SincNet+LIM~\cite{ravanelli2018mutualinfo4spkveri} & 5.80 & n/a & $\bigcirc$ \\ \hline
        \makecell[l]{\ac{M-TTS} \\w/ ASR Phn. Align. SR3} & 6.37 & 0.556 & \\ \hline
        \makecell[l]{\ac{M-TTS} + SpkID loss\\w/ ASR Phn. Align. SR3\\} & \textbf{4.11} & \textbf{0.416} & $\bigcirc$ \\
        \bottomrule
    \end{tabular}}
    \label{tab:Voxceleb1_cmpr_existVSproposed}
    \vspace{-0.1in}
\end{table}

\section{Conclusion}
In this work, \ac{M-TTS} systems including a speaker encoder were used to learn speaker embeddings for the \ac{SV} task. To train speaker embeddings with this method on a dataset without transcripts, 
we compared using manual transcripts, ASR transcripts from E2E and hybrid systems, and phone alignments predicted from the hybrid \ac{ASR}, as TTS text inputs.
We observed that phone alignments performed better than ASR transcripts.
Compared to generative i-vectors and discriminative \ac{ResNet}-\ac{LDE}, the proposed supervised TTS model using only speaker labels achieved better performance. Regarding unsupervised systems, our unsupervised TTS outperformed the i-vector model with 1024 Gaussians but not with regard to the larger version with 2048 Gaussians. How to improve the unsupervised version is the object of further investigation.

A handicap of the proposed model is the high computing cost of training TTS models, roughly 10$\times$ higher than a pure discriminative model.
One way to accelerate computation is to increase the reduction factor in \ac{TTS} training. Another possible future direction is to study an in-depth relationship between the quality of the synthesized speech generated from \ac{M-TTS} and the \ac{SV} performance using the speaker encoder of the \ac{M-TTS}. Finally, using one part of an utterance to extract the speaker embedding while using another part for reconstruction in \ac{M-TTS} training can further improve the embedding by disentangling speaker and phone information~\cite{stafylakis2019ssspkemb}. We intend to explore these directions in future work.



\bibliographystyle{IEEEtran}
\bibliography{main}

\end{document}